\documentclass{ws-ijmpcs}

\usepackage{color}

\begin{document}

\def\gtorder{\mathrel{\raise.3ex\hbox{$>$}\mkern-14mu
             \lower0.6ex\hbox{$\sim$}}}
\def\ltorder{\mathrel{\raise.3ex\hbox{$<$}\mkern-14mu
             \lower0.6ex\hbox{$\sim$}}}

\markboth{Martin Bucher}
{Status of CMB observations in 2015}

%
\catchline{}{}{}{}{}
%

\title{Status of CMB observations in 2015
\footnote{Based on talk delivered in October 2015
at the CosPA 12th International Symposium on Cosmology and Particle
Astrophysics held in Daejeon, South Korea.}
}

\author{Martin Bucher\footnote{bucher@apc.univ-paris7.fr}}

\address{Laboratoire Astroparticules et Cosmologie (APC), 
Universit\'e Paris 7/CNRS\\ B\^atiment Condorcet,
Case 7020, 75205 Paris Cedex 13, France}

\address{Astrophysics and Cosmology Research Unit\\
School of Mathematics, Statistics, and Computer Science \\
University of KwaZulu-Natal,
Durban 4041, South Africa}

\maketitle

\begin{history}
\received{Day Month Year}
\revised{Day Month Year}
\published{Day Month Year}
\end{history}

\begin{abstract}
The 2.725 K cosmic microwave background has played a key role in the 
development of modern cosmology by providing a solid observational foundation 
for constraining possible theories of what happened at very large redshifts and 
theoretical speculation reaching back almost to the would-be big bang 
initial singularity. After recounting some of the lesser known history of this 
area, I summarize the current observational situation and also discuss some 
exciting challenges that lie ahead: the search for B modes, the precision 
mapping of the CMB gravitational lensing potential, and the ultra-precise 
characterization of the CMB frequency spectrum, which would allow the exploitation of 
spectral distortions to probe new physics.
\end{abstract}

\ccode{PACS numbers:}

\section{Introduction}

The Cosmic Microwave Background (CMB) is an almost perfect, nearly isotropic 
blackbody radiation now at a temperature of 2.725 K emanating from the big bang 
itself.  As a practical matter, it is the oldest fossil remnant of the 
primordial universe available to us for precision characterization. The precise 
mapping of the CMB anisotropies in both temperature and polarization has 
allowed us to reconstruct the initial conditions of the universe. As we have 
learned, these initial conditions are remarkably simpler than one might have 
imagined. Twenty years ago people wanting to model how the recent universe 
evolved from initial conditions were forced to insert in their codes 
initial conditions that were at best tentative guesses, but today we believe that we 
know what the correct initial conditions were, with uncertainties in the 
parameters presently at about the few percent level. Today the greatest 
uncertainty in such studies of how the late universe evolved out of these initial 
conditions arises from the modeling of the late time physics itself. While the 
physics of the CMB is linear and simple, the late time physics of the universe 
is nonlinear and complex. Processes---such as gravitational clustering, 
hydrodynamics, magnetogenesis, MHD, star formation, supernova explosions and 
their aftermath, to give an incomplete list---must be modeled, either by starting 
from first principles or in many cases by using more {\it ad hoc} models and relying 
on late time observational data to fix the parameters of these phenomenological 
models. The CMB by contrast offers a relatively clean probe of the state of the 
universe before nonlinear physics kicked in. Even the quantum field theory that 
enters into computing the generation of the primordial density during a prior 
epoch of inflation is conceptually relatively simple: it is, for the most part, 
free field theory, which some quantum field theorists would regard as 
classical, because it is tree level, or order $\hbar ^0$ in the loop expansion. 
In other words, one takes the modes of the classical field theory, normalizing 
them according to classical canonical commutation relations, and one replaces 
the $c$-number coefficients with annihilation and creation operators, just as 
for the simple harmonic oscillator. The study in the 1970s and 1980s of 
semi-classical quantum field theory in a curved spacetime (see for example the 
book by Birrell and Davies\cite{birell} and references therein) raised some 
interesting new issues and became one of 
the principal theoretical tools for the 
subsequent calculation in the mid-1980s of the perturbations predicted from the 
leading order quantum corrections to the classical inflationary cosmology.

This program requires some courage. We do not know the correct theory of 
quantum gravity, but we linearize the classical theory and proceed to quantize 
this linearized theory making predictions as if we had a theory of quantum gravity.  
Perhaps one of the virtues of inflationary cosmology as developed in the 1980s 
is that it protects us from our ignorance. Inflation provides an ingenious 
mechanism for hiding any clues as to what happened during a possible earlier 
Planck epoch, presumably governed by strongly coupled gravity at the quantum level, for 
which we today have hardly any idea as to the character of the underlying 
theory. To be sure, we have superstring theory, which is very promising at the 
formal level. But we do not know how to calculate predictions relevant for the 
very early universe.

Cosmic inflation is an incomplete theory. Some call it the ``inflationary 
paradigm," but I do not like this term, which alludes to Thomas Kuhn's 
``paradigm shifts," which are momentous and revolutionary discoveries after 
which science emerges not the same as before. However the older meaning of 
``paradigm," referring to Latin grammar, is less glorious. According to 
Webster's Third International Dictionary,\cite{webster} 
a paradigm is ``an example of a 
conjugation or declension showing a word in all its inflectional forms." 
Unfortunately, the number of inflationary potentials that have 
been proposed in the literature is immense. The problem is that the theory does 
not make a definite prediction for the form of the inflationary potential, so 
there are, at least technically speaking, almost an infinite number of free 
parameters, corresponding to a free function. We can, however, make predictions 
by assuming the absence of `features,' or bumps and dips in the potential. 
Assuming that the inflationary potential is relatively smooth and well behaved, 
we can, for example, expand the inflationary potential as a power series within 
the observable region where the cosmological perturbations visible to us today 
were imprinted, and retaining only the first few terms, we obtain approximate 
predictions. But there is something unsatisfying about the lack of an 
unambiguous base theory. We can speak of the emergence of a Standard Model of 
Cosmology, but the analogy to the standard electroweak model of Weinberg and 
Salam is imperfect. In that case, there are precisely 26 undetermined 
parameters,\footnote{The 26 standard model parameters are enumerated as 
follows: 3 gauge couplings, (i.e., $g_{color},$ $g_{SU(2)},$ and $g_{U(1)}$), 2 
Higgs potential coupling constants, 3 charged lepton masses, 6 quark masses, 3 
neutrino masses, 4 parameters for the CKM quark mixing matrix and 4 more for 
the neutrino mixing matrix, and finally $\theta _{QCD}.$} and when we talk 
about ``Physics beyond the Standard Model," we know exactly what we are talking 
about.

For my talk I was asked to provide an overview of the observations of the 
cosmic microwave background (CMB) including its history, the present state of 
our knowledge of the CMB, and also some indications about the future of this 
field with a description of the experiments now in progress or being planned, 
as well as 
describing some of the challenges that must be overcome in order to make the 
observations that we would like to make. This is a broad remit, and what is 
included here constitutes an arbitrary selection, reflecting my own 
personal biases. For a more detailed and complete overview, the reader is 
invited to consult my review article,\cite{bucherReview} and in particular the 
many references cited therein.

\section{Pre-History of the CMB}

Here I will try to emphasize some lesser known aspects of the story of the CMB, 
which has become part of the canon of modern cosmology, referring the reader 
elsewhere for a more complete account.\cite{peeblesBook} The history of the CMB 
reaches back much earlier than most people realize, with key observations 
pre-dating the the prophetic set of papers around 1948 of which the most cited 
is Alpher, Bethe, and Gamow,\cite{alphaBetaGamma} in which a CMB temperature of 
approximately 5 K was predicted.\footnote{ For a nice short account of these 
early papers leading to our modern understanding of primordial nucleosynthesis, 
see P.J.E. Peebles, ``Discovery of the Hot Big Bang: What happened in 1948," 
(arXiv:1310.2146)}

The observational evidence for the CMB with a measurement of its temperature 
already existed in studies of the relative intensities of absorption lines in 
spectra of nearby hot stars taken at optical frequencies. Although the 
primary optical transitions are electronic, the electronic levels are split into 
both vibrational and rotational sublevels. The relative populations of these 
sublevels can under the right conditions serve as molecular thermometers that 
can be used to measure the thermodynamic brightness temperature of the 
radiation field at the frequency corresponding to rotational transitions 
between the sublevels. In the general case, this procedure does not work, because 
several competing effects enter into determining the relative level populations 
(e.g., collisions, excitation by UV photons, etc.), each process being 
characterized by its own effective temperature. However, to the extent that the 
coupling to the radiation field dominates over other effects, the 
relative level populations provide an accurate measure of the radiation field 
temperature.

The three molecules originally studied were the diatomic molecules $CN,$ $CH,$ and 
$CH^+.$ The expectation based on laboratory experiments carried out at 
room temperature was that a whole series of `bands' resulting from 
rotational-vibrational splittings into sublevels
should be visible for all these molecules.  But instead only
two sublevels were observed for the CN molecule, corresponding to 
the rotational $J\!=\!1$--$J\!=\!0$ splitting of the ground state, 
and for $CH,$ and $CH^+$ only a single line was observed, suggesting that all 
these molecule are in their rotational ground state. The $J\!=\!1$--$J\!=\!0$ 
rotational transitions for the $CN,$ $CH,$ and $CH^+$ molecules correspond to 
the frequencies 228 GHz, 535 GHz, and 834 GHz (or the wavelengths 1.32 mm 0.56 
mm, and 0.36 mm), respectively.  For comparison, the modern value of the CMB 
temperature converted to a frequency is $\nu _{CMB}=k_BT_{CMB}/h=57$ GHz. For 
the CN relative populations, McKellar\cite{mckellar} deduced an effective 
temperature of around 2.1 K; however, he did not go further to analyze the possible 
consequences of his finding. It was only after the discovery by Penzias and 
Wilson that the coincidence of their direct microwave temperature measurement
with the temperatures recorded using these molecular thermometers was noted. 
It was then recognized 
that the CMB could have been discovered earlier if these measurements had been 
correctly interpreted.

Having described the first observation of the CMB, we now go on to discuss the 
first theoretical work by Alpher, Bethe, and Gamow (although Bethe's name was 
added only to complete the pun). It should be noted that the work by Alpher, 
Bethe, Gamow, and Hermann received little attention at the time, partially for reasons 
that in retrospect are not unreasonable. Today `primordial nucleosynthesis' is 
regarded as one of the `observational pillars' of the hot big bang model, but the 
observational data needed to arrive deductively 
at this conclusion did not exist at the time. In 
1948 the challenge was to explain the abundances of all the elements seen in 
the present universe, ranging from hydrogen to iron and beyond, and primordial 
nucleosynthesis predicted that the baryons in the universe would end up almost 
exclusively in the form of ${}^1$H and ${}^4$He, with literally trace amounts 
of a handful of other light elements: ${}^2$D, ${}^3$He, ${}^7$Li, and ${}^7$Be. 
Around 1948 most researchers looked toward processes within stars for producing the heavier 
elements. They were not aware that stars alone could not account for the entire 
observed helium abundance and that something more than nuclear burning in stars 
was required to explain the observed relative isotope abundances.

It is ironic that in the 1990s before the hot big bang cosmology had become 
thoroughly incorporated into the received wisdom, some critics cited as one of 
their arguments against the big bang the lack of agreement of the 1948 
Alpher, Gamow, and Hermann prediction of a 5 K CMB temperature\cite{alphaBetaGamma} with 
the modern observed value of 2.725 K. Yet if we examine this prediction in 
light of the knowledge at the time and redo the theoretical analysis more 
carefully, we arrive at the conclusion that Alpher et al. were lucky 
not to have been off by an even greater factor. It turns out that unless 
one does extremely sensitive measurements, which have been carried out in the 
meantime, the basic prediction that about a quarter of the baryonic mass is in 
the form of ${}^4$He results over an extremely broad range of values for the 
input parameter $\eta _B$ (the baryon-to-photon number density ratio), which is 
the only adjustable free parameter of primordial nucleosynthesis.
At the time, attention was not so much focused on the possibility of primordial 
nucleosynthesis but more on the nuclear processes in stars,
especially for generating the heavier elements.
It was not then evident that both primordial and stellar
nucleosynthesis were required to explain the current 
observed relative element abundances.

\section{From Penzias and Wilson to COBE}

In many accounts in the secondary literature of the discovery of the CMB by 
Penzias and Wison in 1965 the word `serendipitous' (not exactly among
the most commonly used English words) pops up over and over. While there is 
some truth in this description, in reality the discovery of the CMB was 
less accidental than sometimes depicted. The frontiers of electronics and 
communication have greatly shifted since the mid-1960s. Low noise and high 
frequency had quite a different meaning then, and the problem Penzias and 
Wilson had set out to explore was the ultimate limits of noise for communication in 
the frequency range around 4 GHz. They embarked on the systematic study of this 
question in an effort spanning many years and exploring many possible sources. 
The account depicting of Dicke's Princeton experiment 
on the verge of discovering the 
vestiges of the big bang but being scooped by some folks at Bell Labs who did 
not really know what they were doing does not do justice to 
the contribution of Penzias and Wilson.
At nearly the same time, Robert Dicke and collaborators at Princeton,
unaware of the prior work by Gamow and collaborators, had speculated
that there should be a CMBR and had constructed an apparatus to
measure its signal. However before they could complete their measurement,
the news arrived that Penzias and Wilson had detected the signal
that they had set out to seek, and the outcome was that two papers
were published back to back in the Astrophysical Journal, one
by Penzias and Wilson, preceded by a paper by Dicke, Peebles, Roll
and Wilkinson explaining how this signal could be explained in a hot Big
Bang cosmological model.

There is a long gap between the discovery of Penzias and Wilson of the CMB at 
zeroth order and the discovery of the CMB anisotropy by the COBE DMR in 1992, which was 
accompanied by what remains to date the best measurement of the CMB 
frequency spectrum by the COBE FIRAS instrument. This long gap of almost thirty 
years can largely be understood by the fact that the technology to make the 
necessary measurements was lacking and not that the importance of looking for 
anisotropies and deviations from a blackbody spectrum had not been recognized.
A detailed history of the pre-COBE experiments searching for a possible 
anisotropy in the CMB temperature and for deviations from the blackbody 
form for the frequency spectrum can be found in the reviews by 
Weiss\cite{weiss} and by Readhead and Lawrence.\cite{readhead} The technical 
challenges to be overcome to carry out measurements at the 
required accuracy were formidable, and it took quite some time for the 
upper limits to reach the level of the actual CMB anisotropy as finally 
discovered by COBE.

In 1992 the COBE team announced the first discovery of an anisotropy of 
the cosmic microwave background using the DMR (Differential Microwave 
Radiometer) instrument.\cite{smoot} The DMR consisted of pairs of horns 
pointing in directions separated by approximately $60^\circ $ in the 
sky. The detectors were rapidly switched between the horns and only the 
difference in signal was recorded. Through the spinning and precession 
of the satellite, a data stream consisting of differences in the CMB 
temperature was recorded at three frequencies and sky maps were 
constructed using this data. With the data smoothed to $10^\circ ,$ the 
COBE team found an rms anisotropy of $30\pm 5 \mu K$ (or $\Delta 
T/T=11\times 10^{-6})$ for the primordial blackbody signal. The shape of 
the spectrum was found consistent with scale invariance (although 
with large uncertainties, i.e., $n=1.1\pm 0.5$) as predicted by 
inflationary models. The COBE FIRAS instrument probed the absolute 
frequency spectrum of the CMB by rapidly switching a Fourier transform 
spectrometer between the sky and an artificially constructed blackbody
source. COBE FIRAS constrained the rms deviations of the CMB from a 
perfect blackbody to be less than 5 parts in $10^5$ over the frequency 
range 60 -- 630 GHz.\cite{firas}

Competing theoretical cosmological models had an undetermined or 
floating normalization, but when normalized to COBE, a prediction for 
the normalization of the primordial power spectrum of density 
fluctuations followed, usually expressed in terms of $\sigma _8$ (i.e., 
the rms fractional mass fluctuation 
$\delta M/M$ within a sphere of radius $8 h^{-1}Mpc$), 
could be obtained, and this predicted $\sigma _8$ could be compared 
to data from galaxy surveys. Models with adiabatic fluctuations, as 
produced in the simplest inflationary models, produced higher values of 
$\sigma _8$ than models with isocurvature perturbations or models with 
structure formation through topological defects.

The next step was to explore how the CMB angular power spectrum extended 
to smaller angles. A host of ground and balloon based experiments 
followed on the COBE detection, and it took some time for the observational 
situation to become clear. For a number of years, a plot summarizing the 
observational situation showing all the experiments extending the COBE 
DMR experiment to smaller angular scales somewhat resembled a target 
where the marksman often altogether missed the target.
Theorists could hand pick the points they liked, claiming that based on 
inside information certain points should be believed more than others. 
(See for example the figure of Max Tegmark in ref. \refcite{turner}.)
The observations improved gradually.
It is usually the Boomerang \cite{boomerang}
and Maxima \cite{maxima} experiments that are credited with having provided the first 
clear and convincing evidence of the acoustic oscillations.
The next major step was the WMAP space mission, often described as the 
``second-generation" CMB space mission, following on COBE. 

\section{The WMAP and Planck space missions}

The NASA WMAP satellite was launched in 2001 and delivered its first results
in 2003 concerning the CMB temperature anisotropy over the whole sky. WMAP's
first CMB polarization results were reported in 2006, and WMAP continued taking
data until 2010. Periodic updates of the WMAP results were subsequently published taking 
into account more data and refining the analysis. Like COBE, WMAP surveyed the
entire sky, but because WMAP used a pair of telescopes rather than horns pointing
directly at the sky, WMAP had a superior angular resolution.  
WMAP observed in five frequencies: 23, 33, 41, 61, and 94 GHz.
The lowest frequency bands were the most contaminated by Galactic synchrotron radiation,
and the highest (or highest two) frequency band was the cleanest. Because of
the HEMT (high electron mobility transistor) (i.e., coherent amplification) 
detection technology employed, WMAP could not reach to higher frequencies
were Galactic thermal dust contamination starts to become dominant.

I vividly recall the first press conference when the WMAP 
collaboration announced their first scientific results, and given that this was 
the first post-COBE CMB space mission, there were great expectations that some 
bombshell was going to be dropped. Boomerang\cite{boomerang} and 
Maxima\cite{maxima} had presented fairly convincing evidence for the first 
Doppler peak at precisely the position predicted by $\Lambda $CDM combined with 
simple inflationary models. Even though these observations killed off a lot of 
models---or at least made most of these models very difficult to defend, the 
error bars were still quite large, and there remained the possibility that WMAP 
with its high precision and full sky coverage would conclude ``none of the 
above'' for the many still contending theoretical models. As a theorist, this 
was certainly my hope, because simple models of single scalar field inflation 
are rather boring. At the press conference, John Bahcall summed up that ``the 
greatest surprise [from WMAP] was that there was no surprise." Indeed WMAP 
found that their measured temperature power spectrum was consistent with the 
simplest six-parameter model. The six parameters were $A_S,$ $n_S,$ $H_0,$ 
$\omega _b=h^2\Omega _b,$ $\omega _c=h^2\Omega _c,$ and $\tau .$ Or 
equivalently, one could use any other six independent parameters derived from 
these parameters.

One of the interesting novelties from among the WMAP results was the 
measurement of the spectral index of the power spectrum, which prior to WMAP 
was consistent with $n_S=1,$ which would be the prediction if there were an 
exact unbroken scale symmetry. While inflationary potentials can be 
constructed---or some might say `designed'---through fine tuning to give $n_S$ 
arbitrarily close to one, generically inflation predicts some deviation from 
exact scale invariance, most likely but not inexorably on the side $n_S<1.$ The 
three-year WMAP results in the framework of the previously described 
six-parameter model measured $n_S=0.958\pm 0.016,$ thus excluding a 
scale-invariant primordial power spectrum at approximately $2.6\sigma .$ For 
many this result strengthened the case for inflation, as opposed to some 
unspecified more symmetric theory for imprinting the primordial cosmological 
perturbations.

WMAP also was able to place interesting constraints on primordial 
non-Gaussianity and discovered a few `anomalies' at modest statistical 
significance, a topic to which we shall return below in our discussion of Planck. 
Until the Planck 2013 Cosmology Results were announced, the WMAP 
characterization of the temperature power spectrum on intermediate scales 
remained unrivaled, although ground-based experiments, most notably 
ACT\cite{actPSref} and SPT,\cite{sptPSref} succeeded in extending the 
measurement of $C^{TT}_\ell $ to smaller scales extending to around $\ell 
\approx 10^4.$

In these proceedings I shall not dwell on the ESA Planck space mission because 
Planck has over the last few years released two sets of cosmology results, one 
in 2013 and another in 2015. These results have been the subject of many plenary 
conference presentations. It is therefore doubtful that I can in the small 
number of pages available tell you something that you have not already heard or 
read elsewhere. For this reason I invite you the consult the relevant 2013 and 
2015 papers for a complete account. Let me however give you my own take on 
the main scientific impact of the Planck.

Planck had the capability to improve on WMAP in several ways owing to its ten 
times better sensitivity and two times better angular 
resolution.\cite{planckBluebook} The dimensions of the Planck and WMAP 
telescopes were roughly comparable, with a usable diameter of slightly over a 
meter, the main difference between the two missions being the types of 
detectors deployed. WMAP deployed coherent detectors passively cooled to about 
95 K and less sensitive, while Planck used, at least for the 
six highest HFI (High Frequency Instrument) frequency channels, more sensitive bolometers 
that were cryogenically cooled to $\approx 100 $ mK. On a space mission, 
bolometers are more risky because failure of the cooling system to perform 
adequately would prevent any useful data from being taken. Bolometers however 
present two significant advantages: (1) Bolometers function nearly at the limit 
of the inherent quantum noise of the incident photon flux, whereas the 
sensitivity of coherent detectors is worse than this fundamental limit by a 
substantial factor, and (2) coherent detectors for measuring CMB anisotropies 
are able to operate only up to about 100 GHz, whereas bolometers function 
successfully at much higher frequencies extending into the Wien tail of the CMB 
blackbody frequency spectrum. WMAP mapped the microwave sky in five bands 
centered at 23, 33, 41, 61, and 94 GHz, whereas Planck mapped the sky in nine 
bands centered at 30, 44, 70, 100, 143, 217, 353, 545, and 857 GHz. The extended 
reach of Planck thanks to these higher frequency bands allowed measurements 
where the dust contributes substantially to the sky signal---and even 
dominates over the CMB in the highest bands, thus allowing better cleaning of 
nonprimordial contaminants and substantially reducing uncertainties in the 
possible role of dust contamination in the final cleaned CMB maps. The highest 
Planck frequency (at 857 GHz) is situated far into the Wien tail of the CMB, so 
that except very near the galactic poles, one sees hardly any CMB. Almost all 
the signal comes from thermal dust emission from our Galaxy as well as from 
infrared galaxies. The presence of these higher frequency channels constitutes
an additional 
advantage. Since CMB measurements are almost always diffraction limited, given 
the roughly comparable dimensions of the telescopes, Planck benefits from a 
better angular resolution.

\begin{figure}
\begin{center}
\includegraphics[width=12cm]{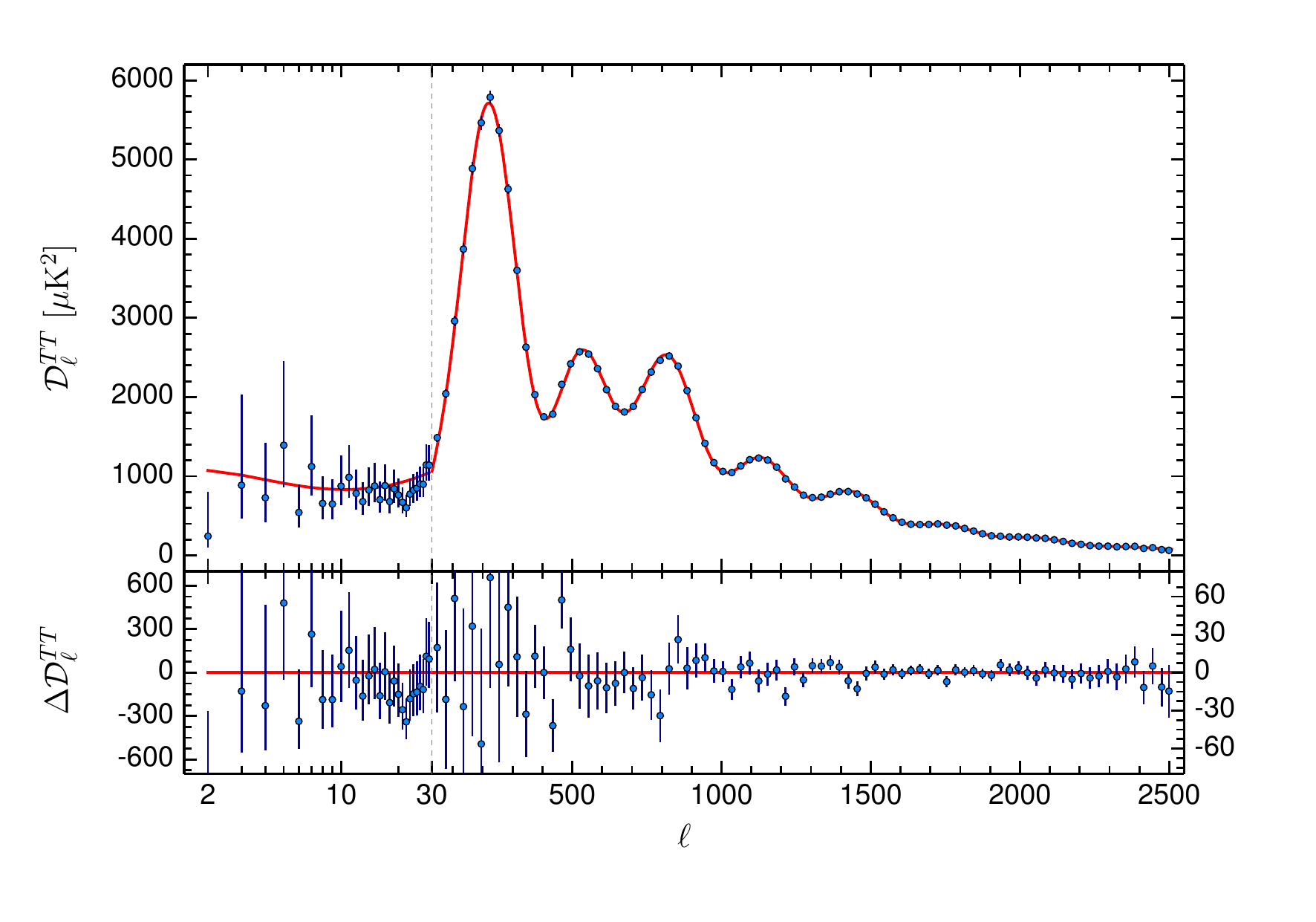}
\end{center}
\vskip -30pt 
\caption{{\bf Planck 2015 TT Power Spectrum.} 
The upper subplot shows the binned $C^{TT}_\ell $ power spectrum
(where $D_\ell ^{TT}=\ell (\ell+1)C^{TT}_\ell /2\pi $) compared
to the best fit six-parameter theoretical model (red), and the
lower plot shows a zoom of the residuals. The fit is quite good
except for a dip around $\ell \approx 20,$ whose statistical
significance is around $2\sigma .$
{\it (Credit: ESA/ Planck Collaboration)}
}
\label{PlanckTTFig}
\end{figure}

The temperature power spectrum reported in the Planck 2013 Cosmological Results 
release was consistent with the same six-parameter model used for the WMAP 
analysis. Figure \ref{PlanckTTFig} shows the temperature power spectrum from 
the 2015 Cosmology release. (For more details, see 
Refs.~\refcite{Planck2015Overview} and \refcite{planck2015params}.) One could 
say that the John Bahcall's 2001 WMAP summary that the ``greatest surprise was 
that there was no surprise'' could equally well be applied to the Planck 
results, despite the considerable scope to detect anomalies as explained above. 
The Planck $C^{TT}_\ell $ power spectrum shows several improvements of a qualitative 
nature relative to the WMAP $C^{TT}_\ell $ power spectrum. While only three acoustic 
oscillations are clearly visible in the WMAP power spectrum before the error 
bars start to blow up at higher $\ell ,$ the Planck power spectrum shows clearly five 
acoustic peaks, and moreover the decay of the damping tail is precisely mapped 
out to about $\ell \approx 2500$ where things start to fall apart. At this 
point beam smearing combined with the limited sensitivity causes the error bars 
to almost literally blow up toward higher $\ell ,$ so that beyond $\ell \approx 
2500$ there is little useful data.

The other result from 2013 that was perhaps surprising to a lot of theorists 
was the stringent upper limits established by Planck on the possible primordial 
non-Gaussianity. In the WMAP data, there were some hints at very modest 
statistical significance of an $f_{NL}>0.$ Yadav and Wandelt\cite{wandelt} 
reported that according to their independent analysis of the WMAP data, 
$f_{NL} \in [27,147]$ at 99.5\%  confidence level 
with a central value of $f_{NL}=87.$ Other analyses 
found somewhat lower values, but despite disagreement on the details, most 
analyses found a hint of a signal with $f_{NL}>0$ at low statistical 
significance. In the wake of these developments, theorists worked feverishly to 
devise models predicting a 
large primordial bispectral non-Gaussianity. Many of these 
theorists anticipated a Planck confirmation of the hints found by WMAP, but in its 
2013 analysis based on the temperature maps, the Planck Collaboration 
reported $f_{NL}=2.7\pm 5.8,$ thus laying to rest the hint of a detection 
that some had seen in the WMAP data. The 2015 Planck results, which 
additionally included polarization, gave consistent constraints.

\begin{figure}
\begin{center}
\includegraphics[width=6cm]{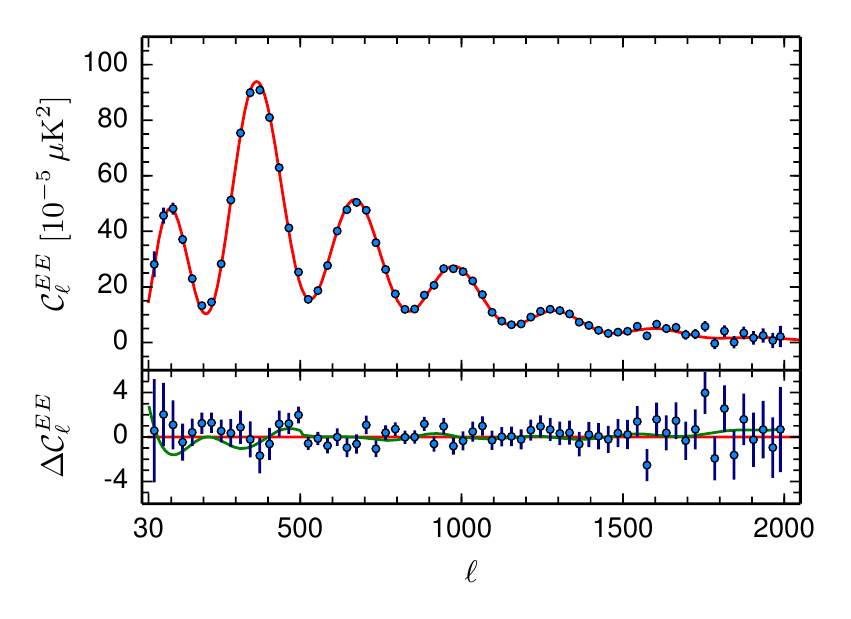}
\includegraphics[width=6cm]{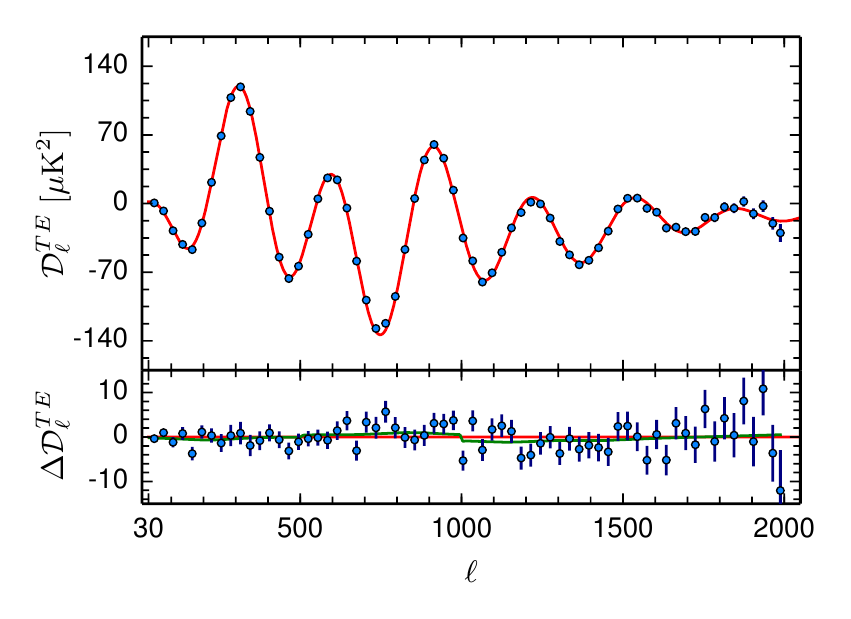}
\end{center}
\vskip -20pt
\caption{{\bf Planck polarization power spectra.} 
The $C^{EE}_\ell $ (left) and $C^{TE}_\ell $ (right) power spectra reported in 
the Planck 2015 release show that the observed polarized anisotropies are 
broadly consistent with the six-parameter concordance model. The error bars 
shown should be understood as unreliable and as a lower bound on the actual 
errors, because as the Planck 2015 papers reported, there remains some evidence 
of unaccounted or uncorrected systematics in the polarization maps, most 
notably $T\to E$ leakage.
{\it (Credit: ESA/ Planck Collaboration)}
}
\label{PlanckPolarizationFigure}
\end{figure}

The Planck 2015 results also included the CMB $C^{EE}_\ell $ auto-correlation 
power spectrum as well as the $C^{TE}_\ell $ cross-correlation power spectrum. 
(See Fig.~\ref{PlanckPolarizationFigure}.) The polarization of the CMB had 
first been discovered by DASI in 2002\cite{dasi} and was later mapped over the 
whole sky by WMAP at lower resolution and sensitivity than in the 2015 Planck 
results. The first WMAP polarization results were released in 2005 based on the 
WMAP 3-year data, and updated using the WMAP 5-year, 7-year, and 9-year data 
were also released. 
When analyzed independently without using the temperature data,
the Planck polarization power spectrum gives cosmological parameters 
consistent with those obtained using the TT data alone and with those using
all the data.
This fact indicates that the CMB polarization tells a consistent 
story, corroborating the story told by the temperature anisotropies alone. The 
signal-to-noise ratio of the Planck $a_{\ell m}^E$'s hovers near or below unity until 
about $\ell \approx 1000,$ beyond where the error bars start to blow up.

The Planck team also studied the various anomalies found by the WMAP team at 
modest statistical significance. These include bipolar disorder (or dipole 
modulation of the power spectrum), alignments of certain low $\ell $ multipole 
moments, and the `cold spot,' all of which are results at low statistical 
significance. This means that one is free either to reject them as statistical 
flukes, or to take them as interesting hints of new physics, following one's own 
personal tastes. Some believed that Planck with its higher frequency channels 
might make these anomalies go away by showing that they resulted from 
underestimating or improperly correcting for Galactic 
dust contamination. This however did not happen. The anomalies seen by WMAP were 
broadly confirmed by Planck at comparable statistical significance. This
result is not at all surprising because the limiting factor for both 
experiments is cosmic variance. Unfortunately, future measurements of the CMB 
temperature cannot hope to shed additional light on this question.
They too will be seeing that same sky and will be limited by exactly the same 
cosmic variance. There is however room for modest improvement resulting from 
using polarization, perhaps from the late-2016 Planck low $\ell $ polarization 
results when these become available.

In early 2015 the Planck Collaboration 
released its first cosmological results including the  
polarization data. The measurements of the EE polarization power spectrum at 
$\ell >30$ is consistent with the 2013 results using TT alone, and perhaps more 
importantly, the cosmological parameters found using EE alone tell a story 
consistent with the TT only results, 
showing that the measured polarization power spectrum 
corroborates the story told by TT. In the 2015 cosmology release, however, it 
was noted that the polarization data contained some evidence of residual
systematic errors and that 
the polarization error bars given are not fully trustworthy. Moreover, the low 
$\ell $ polarization power spectrum (for $\ell <30$) had not been released 
because the analysis was still ongoing. In late 2016 the Planck Collaboration 
intends 
to release a new set of results that will, most importantly, provide Planck 
measurements of the polarization extending all the way down to $\ell =2.$ This 
data will help pin down the reionization optical depth as well as giving Planck 
constraints on $r$ from the reionization bump. It is however doubtful that these 
constraints will be better than the most recent constraints established by 
BICEP2/Keck array team.

We now turn to discussing the B modes of the CMB polarization, which have received 
a lot of attention, especially since the March 2015 announcement by the BICEP2 
team of a purported discovery of B modes using measurements in just one 
frequency channel at 150 GHz.\cite{Bicep2Claim} Based on CMB polarization 
measurements in a patch covering only about 2\% of 
the sky chosen for its low level of 
galactic foregrounds as measured using temperature maps, the BICEP2 team found 
a B mode amplitude that if entirely primordial would correspond to 
$r=0.20^{+0.07}_{-0.05}$ with $r=0$ disfavored at $7\sigma .$ According to the 
interpretation put forth by the BICEP2 team, an explanation based on Galactic 
foregrounds could be excluded because of the extremely large magnitude of the 
observed signal. This claim made the front pages of newspapers around the world 
and sparked great enthusiasm in the popular science press. BICEP2 claimed to 
have investigated the predictions of six dust models and to have found that 
none of these models could 
account for dust contamination having such a large magnitude. There was only a slight 
problem: there was no publicly available data to fix the parameters of these 
models, a difficulty not sufficiently appreciated by the four BICEP2 PIs. The 
only data that could settle this question was in the Planck high frequency 
polarized dust maps (particularly the polarized maps at $353$ and $545$
GHz whose signal is almost exclusively thermal
dust emission from our Galaxy), 
but at that time 
the Planck Collaboration had not publicly 
released these maps, nor had they yet presented their analysis. In the early 
summer a paper by Flauger, Spergel, and Hill\cite{flauger} appeared arguing 
that all the data available at that time 
concerning polarized dust emission
implied that all the signal could be 
attributed to Galactic dust contamination, although a primordial signal could 
not be ruled out. In September 2015 the Planck Collaboration finally published 
the results of their polarized dust analysis showing that in fact the polarized 
dust magnitude measured by Planck (primarily using the 353 GHz channel) in the 
BICEP2 field was sufficiently high to account for all the 
signal.\cite{PlanckXXX} Subsequently Planck and BICEP2 agreed to carry out a 
joint cross-correlation study in order to determine the actual level of dust 
contamination. The result of this analysis was that the detection, initially 
claimed at $7\sigma ,$ went away, and a new upper bound on $r=T/S$ with 
$r<0.12$ at 95 \% confidence was established.\cite{planckBicepJoint} Subsequent 
work by the BICEP2/Keck Array team incorporating more recent data improved this 
bound to $r<0.09$ at a 95 \% confidence level.\cite{bicepBetterLimit}

There are at present many ground and balloon based experiments underway, for 
example ACT, BICEP2/Keck Array, CLASS, EBEX, Piper, Polar Bear/Simons Array, 
Quijote, QUBIC, Spider, and SPT, aiming to improve the upper bounds on $r$ or to make 
a first detection. We discuss some of the most ambitious of the initiatives in 
the next two sections.

\section{Where we stand today}

With some risk of oversimplification, the status of the CMB today could be 
summed up as follows. The Planck space mission together with ground based 
experiments such as ACT and SPT extending to smaller scales have mapped the CMB 
temperature anisotropies to reach almost the cosmic variance limit, and at 
present the frontier of CMB research is the search for primordial B modes that 
were presumably generated during cosmic inflation, as well as an ultra-precise 
characterization of the absolute frequency spectrum by improving on COBE FIRAS.

Before going on to discuss the search for B modes, let us nuance this claim 
somewhat. At low multipole number $\ell ,$ the Planck determination of the 
temperature power spectrum is limited by cosmic variance---that is, the fact 
that we can observe only one sky. If we accept the hypothesis that the 
primordial scalar perturbations were imprinted by an isotropic Gaussian 
stochastic process, all useful information in the CMB temperature anisotropies 
can be compressed into the 
measured power spectrum $C_\ell ^{TT},$ and even with the most perfect 
measurement, there always remains a residual rms fractional uncertainty error of 
$\sqrt{2/(2\ell +1)}$ in the measurement of each $C_\ell $ for the underlying 
isotropic statistical process that generated our random sky map. Even with the 
best possible measurements, there is a maximum $\ell $ beyond which very little 
precise information can be obtained regarding the primordial perturbations 
given that beyond $\ell \approx 3000$ other effects rapidly kick in and 
dominate the sky anisotropy signal. Only measurements of the primordial signal averaged 
over broad bands and with large fractional error bars can be obtained beyond 
this approximate threshold.

This transition from a regime at low $\ell $ where the primordial CMB signal 
dominates to a regime at high $\ell $ where the primordial signal becomes a 
sideshow is quite abrupt for a number of reasons. Most of the foregrounds that 
dominate on small angular scales have the power spectrum of white noise or of 
point sources---that is, $C^{TT}_\ell \sim \ell ^0.$ The primordial CMB 
temperature power spectrum, by contrast, at low $\ell $ 
scales very approximately as $\ell ^{-2}$ (if one ignores the acoustic 
oscillations) but at higher $\ell $ falls exponentially due to Silk damping 
or `viscosity.' Point sources can be masked, but there are unsurmountable 
limitations to how much contamination can be mitigated by fancy `component 
separation' techniques because each point source has its own unique frequency 
spectrum.

The end result, as estimated by Fisher forecasts for future CMB space missions, 
is that the error bars on the cosmological parameters may shrink each by a 
factor of approximately 2--3.\footnote{We shall not follow the Dark Energy 
community, which uses a figure of merit consisting of the volume factor by which 
the allowed parameter space is shrunk as the result of better data. This is a 
dangerous accounting practice susceptible to manipulation, because for example 
a 30\% improvement in the error bars in each of, say, 30 parameters of a model 
can be billed as an improvement by a factor of almost 50,000. But we all know 
that a 30\% improvement in the error bars is always subject to a lot of 
caveats, subjective choices, and modelling uncertainties.} It has been argued 
that one can do better with measuring the EE polarization power spectrum 
because despite the fact that the CMB polarization is harder to measure 
experimentally on account of its smaller absolute magnitude, the non-primordial 
contaminants have a smaller fractional polarization than the primordial CMB, 
and thus for polarization the wall at around $\ell \approx 3000$ moves to 
somewhat higher $\ell .$

The preceding comments should not be misinterpreted as implying that exploring 
the CMB beyond $\ell \gtorder 3000$ is not of great interest. As the papers 
from the ACT and SPT collaborations have shown, there is a lot of interesting
science to be done by observing the microwave temperature and polarization
anisotropy on very small 
angular scales, but the primary science goals are not the primordial scalar 
mode power spectrum but rather the science of what are often termed as the 
``secondary anisotropies.'' Gravitational lensing of the CMB is a very 
interesting (and clean) probe of the inhomogeneities of the distribution of 
matter lying between ourselves and the last scattering surface at $z\approx 
10^3.$ In a certain sense, the idea behind these observations is the same as 
for studies of weak lensing in the optical and infrared bands, where one 
observes the ellipticities of galaxies after smoothing. The simplest model 
would be that except for the lensing shear, the ellipticities would be random and 
uncorrelated, so that after averaging over many galaxies one recovers the lensing field.  
However intrinsic alignments add extra noise and biases to the reconstructed 
lensing field maps. A few differences between CMB and other lensing 
surveys are worth highlighting: (1) The CMB constitutes the most distant 
exploitable source plane, whereas weak lensing of galaxies does not 
extend as far back in redshift. This means that linear theory is more 
applicable to interpreting CMB lensing results than to the weak lensing of 
galaxies. (2) For constructing a lensing map based on the alignments of the 
ellipticities of galaxies, it is assumed that there are no intrinsic 
alignments, which would act to bias measurements. CMB lensing is free of this 
complication and can be used by studying cross correlations to test for 
and to characterize such 
intrinsic alignments. Moreover, combining with CMB lensing also provides more 
powerful `tomography,' in order to try to recover depth information.

Another frontier of CMB research is the measurement of the absolute frequency 
spectrum of the CMB. It is remarkable that the COBE FIRAS experiment of 25 
years ago still remains the best measurement of the absolute frequency 
spectrum, notwithstanding some improvement at low frequencies by the ARCADE
balloon borne experiment.
By contrast, great progress has been made improving on the COBE 
DMR first detection of the primordial CMB anisotropy, and much of this 
improvement derives from ground and balloon based experiments and not 
just the two post-COBE CMB space missions WMAP and Planck. While differential 
measurements can be carried out through the atmosphere, absolute measurements 
require subtractions that are not feasible. Technology has greatly improved 
since FIRAS, and an experiment called PIXIE\cite{pixie} 
has been proposed very roughly 
speaking to redo the FIRAS experiment with polarization and with a sensitivity 
approximately two orders of magnitude better. We also parenthetically note that 
PIXIE also proposes to search for B modes with the same instrument (using a 
substantially different technique).
For an interesting summary of the new science made possible
through measuring the absolute spectrum, see Ref.~\refcite{silkChluba}
and references therein.

\section{Conclusion}

We have seen that observations of the CMB both of its anisotropies in 
temperature and polarization and of its absolute frequency spectrum have played 
a key role in putting modern cosmology on a firm observational footing, so that 
what is now known as the ``standard model of cosmology" or the early universe 
is not just theoretical speculation. We should be careful not to exaggerate the 
successes of this model, because there remain a lot of open questions, and in 
certain respects (e.g., the functional form of the primordial power spectrum), 
the model provides what may be fairly described more as a fitting formula than 
a complete theory. For example, the standard model as proposed by Weinberg and 
Salam, even though theorists may not like it 
so much for various well justified reasons, has a well 
defined number of parameters. There are no free functional forms. Even 
including massive neutrinos, which are now known to exist and often described as 
`physics beyond the standard model,' one must admit that the few extra terms 
and parameters from the theoretical point of view hardly constitute a 
revolutionary update of 
the theoretical physics of the late 1960s and 1970s. Indeed there is room for a 
compelling theoretical model realizing inflation that would provide a 
prediction for the functional form of the inflationary potential, or for an 
alternative to inflation. But this is a subject on which we have seen many 
papers, but unfortunately there have been no qualitative breakthroughs.

To my mind, the most interesting future challenge of CMB research is the search 
for primordial B modes. As explained above, the scalar power spectrum has 
already been characterized close to the ultimate limit imposed by cosmic 
variance. On the other hand, regarding B modes, all that we have for the moment is 
ever improving upper bounds. The last five years have seen a remarkable 
improvement in the quality of these bounds, to the point now that the best 
bounds on $r$ derive from B modes and no longer from the shape of the $TT$ 
power spectrum at $\ell \ltorder 100.$ The bounds derived 
from the shape of the temperature spectrum are highly model dependent, whereas 
the bounds from B modes are extremely robust.

There remains substantial scope for improving on measurements of the B modes. 
The most ambitious projects currently in the planning stage aim to be able to 
make a detection of $r=0.001$ at high statistical significance, and if one is 
very (perhaps unrealistically) optimistic about the unknowns (e.g., complexity 
of foregrounds, ability to overcome systematic uncertainties), one might hope 
to be able to push beyond the goal.

At this point, apart from the experiments already underway, several 
ambitious initiatives are being planned. In the suborbital class, there is the 
ground-based American S4 (``Stage 4") program, whose details remain to be 
defined. S4 aims to deploy of order $\textrm{(few)}\times 10^5$ 
detectors from the ground in order to detect primordial B modes and also map 
the lensing potential from the ground. What can actually be achieved from the 
ground is subject to considerable uncertainty on account of atmospheric interference
and the paucity of bands through which the atmospheric contamination is not overwhelming. 

From space three satellite missions are presently under consideration: 
the Japanese-led LiteBird mission (described in more detail in the contribution 
by Hirokazu Ishino in these proceedings), an ESA M5 proposal whose details 
are not yet defined, and PIXIE.
The ESA M5 proposal could be similar to the previous 
C0rE and COrE+ proposals, or it could take the form of a joint ESA-JAXA-NASA 
mission building on the success of the LiteBird JAXA/NASA joint phase A study 
now underway. Another possibility is a new version of the PIXIE proposal. In 
any case, it is clear that there still remains much exciting new science to be 
done in the area of CMB observation. It is difficult to predict exactly what path this 
field will take.


\section*{Acknowledgments}

MB would like to thank the organizers of the CosPA 2015 12th International 
Symposium on Cosmology and Particle Astrophyscis held from 12-16 October in 
Daejeon, Korea for their hospitality and for an impeccably well organized and 
interesting conference.


\begin{thebibliography}{0}    

\bibitem{bucherReview}
M. Bucher, ``Physics of the cosmic microwave background anisotropy,"
Int. J. Mod. Phys. D24 (2015) 1530004 (arXiv:1501.04288)

\bibitem{webster}
P.B. Gove, Ed.,
{\it Webster's Third New International Dictionary (unabridged),}
(Springfield, MA: Merrian-Webster, 1961)

\bibitem{lerner}
E. Lerner,
{\it The Big Bang Never Happened: A Startling Refutation of the Dominant
Theory of the Origin of the Universe,}
(New York: Times Books, 1991).

\bibitem{birell}
N.D. Birrell and P.C.W. Davies,
{\it Quantum fields in curved space,}
(Cambridge: Cambridge University Press, 1984)

\bibitem{alphaBetaGamma}
R.A. Alpher, H. Bethe, and G. Gamow, ``The origin of chemical elements,"
Phys. Rev. 73 (1948) 803

\bibitem{planck2015params}
Planck Collaboration (P.A.R. Ade et al.),
``Planck 2015 results. XIII. Cosmological parameters
(arXiv:1502.01589)

\bibitem{Planck2015Overview}
Planck Collaboration (R. Adam et al.), 
``Planck 2015 results. I. Overview of products and scientific results,"
(arXiv:1502.01582) 

\bibitem{wandelt}
A.P.S. Yadav and B.D. Wandelt,
``Detection of primordial non-Gaussianity (fNL) in the WMAP 3-year data at above 99.5\% confidence,"
Phys. Rev. Lett. 100 (2008) 181301 (arXiv:0712.1148)

\bibitem{planckNG}
Planck Collaboration (P.A.R. Ade et al.),
``Planck 2013 results. XXIV. Constraints on primordial non-Gaussianity,"
Astron. Astrophys. 571 (2014) A24
(arXiv:1303.5084)

\bibitem{planckLensing}
Planck Collaborations (P.A.R. Ade et al.),
``Planck 2013 results. XVII. Gravitational lensing by large-scale structure,"
Astron. Astrophys. 571 (2014) A17
(arXiv:1303.5077)

\bibitem{wmapFirstYearsCosmoParams}
D.N. Spergel, L. Verde, H.V. Peiris et al.,
``First Year Wilkinson Microwave Anisotropy Probe (WMAP) Observations: 
Determination of Cosmological Parameters,"
Ap. J. Suppl. 148 (2003) 175
(arXiv:astro-ph/0302209)

\bibitem{wmapThreeYearCosmo}
D.N. Spergel et al.,
``Three-Year Wilkinson Microwave Anisotropy Probe (WMAP) 
Observations: Implications for Cosmology,"
Ap. J. Suppl. 170 (2007) 377
(arXiv:astro-ph/0603449)

\bibitem{flauger}
R. Flauger, J.C. Hill and D.N. Spergel,
``Toward an Understanding of Foreground Emission in the BICEP2 Region,"
JCAP 1408 (2014) 039 
(arXiv:1405.7351)

\bibitem{Bicep2Claim}
The BICEP2 Collaboration (P.A.R. Ade et al.),
``BICEP2 2014 I: Detection of B-mode Polarization at Degree Angular Scales by BICEP2,"
Phys. Rev. Lett. 112 (2014) 241101
(arXiv:1403.3985)

\bibitem{PlanckXXX}
Planck Collaboration (R. Adam et al.),
``Planck intermediate results. XXX. The angular power spectrum of polarized 
dust emission at intermediate and high Galactic latitudes,"
Astron. Astrophys. 586 (2016) A133 (arXiv:1409.5738)

\bibitem{planckBicepJoint}
The BICEP2/Keck and Planck Collaborations, 
``A Joint Analysis of BICEP2/Keck Array and Planck Data,"
Phys. Rev. Lett. 114 (2015) 101301
(arXiv:1502.00612)

\bibitem{bicepBetterLimit}
Keck Array, BICEP2 Collaborations (P.A.R. Ade et al.),
``BICEP2/Keck Array VI: Improved Constraints On Cosmology and Foregrounds When 
Adding 95 GHz Data From Keck Array,"
Phys. Rev. Lett. 116 (2016) 031302 
(arXiv:1510.09217)

\bibitem{actPSref}
S. Das, T. Louis, M.R. Nolta et al.,
``The Atacama Cosmology Telescope: Temperature and 
Gravitational Lensing Power Spectrum Measurements from Three Seasons of Data,"
JCAP 04 (2014) 014
(arXiv:1301.1037)

\bibitem{sptPSref}
R. Keisler et al.,
``A Measurement of the Damping Tail of the Cosmic Microwave Background Power Spectrum with 
the South Pole Telescope," Ap. J. 743 (2011) 28 (arXiv:1105.3182)

\bibitem{planckBluebook}
Planck Collaboration,
``Planck: The Scientific Programme," (also known as the Blue Book)
(2005)
http://sci.esa.int/planck/47334-planck-the-scientific-programme/

\bibitem{boomerang}
C. B. Netterfield et al.,
``A Measurement by BOOMERANG of Multiple Peaks in the Angular Power Spectrum 
of the Cosmic Microwave Background,"
Ap. J. 571 (2002) 604
(arXiv:astro-ph/0104460)

\bibitem{maxima}
A.T. Lee et al.,
``A High Spatial Resolution Analysis of the MAXIMA-1 Cosmic Microwave Background Anisotropy Data,"
Ap. J. Lett. 561 (2001) 1
(arXiv:astro-ph/0104459)

\bibitem{dasi}
J. Kovac et al., ``Detection of Polarization in the Cosmic Microwave Background using DASI,"
Nature 420 (2002) 772
(arXiv:astro-ph/0209478)

\bibitem{peeblesBook} 
P.J.E. Peebles, L.A. Page and R.B. Partridge, {\it Finding the Big Bang,}
(Cambridge: Cambridge University Press, 2009)

\bibitem{mckellar} 
A. McKellar,    
``Evidence for the Molecular Origin of Some Hitherto Unidentified Interstellar Lines,"
PASP 52, (1940) 187

\bibitem{pixie}
A. Kogut, D.J. Fixsen et al.,
``The Primordial Inflation Explorer (PIXIE): a nulling polarimeter for 
cosmic microwave background observations," JCAP 2011 (2011) 025 (arXiv:1105.2044) 

\bibitem{firas}
D.J. Fixsen et al., ``The Cosmic Microwave Background Spectrum 
from the Full COBE FIRAS Data Set," Ap. J. 473 (1996) 576

\bibitem{smoot}
G.F. Smoot et al., ``Structure in the COBE differential 
microwave radiometer first-year maps,"
Ap. J.  396 (1992) L1

\bibitem{silkChluba}
J. Silk and J. Chluba, ``Next Steps for Cosmology,"
Science  344 (2014) 586

\bibitem{liteBird}
T. Matsumura et al.  ``Mission design of LiteBIRD,"
J. Low Temp. Phys.  176 (2014) 733 (arXiv:1311.2847)

\bibitem{stageFourOne}
K.N. Abazajian et al.,
``Inflation Physics from the Cosmic Microwave Background and Large Scale Structure,"
Astropart. Phys. 63 (2015) 55
(arXiv:1309.5381)

\bibitem{stageFourTwo}
K.N. Abazajian et al.,
``Neutrino Physics from the Cosmic Microwave Background and Large Scale Structure,"
Astropart. Phys. 63 (2015) 66 (arXiv:1309.5383)

\bibitem{pixie}
A. Kogut, D.J. Fixsen et al.,
``The Primordial Inflation Explorer (PIXIE): A Nulling Polarimeter for Cosmic 
Microwave Background Observations,"
JCAP 07 (2011) 025 (arXiv:1105.2044)

\bibitem{turner}
M. Turner, ``Cosmology Solved?," (arXiv:astro-ph/9811447)

\bibitem{weiss}
R. Weiss, ``Measurements of the Cosmic Microwave Background Radiation,"
Ann. Rev. Astron. Astrophys. 18 (1982) 489

\bibitem{readhead}
A.C.S. Readhead and C.R. Lawrence, ``Observations of the Isotropy of the 
Cosmic Microwave Background Radiation," Ann. Rev. Astron. Astrophys. 30 (1992) 653

\end{thebibliography}
\end{document}